\documentstyle[12pt,aaspp4]{article}
\lefthead{JENKINS ET AL.}
\righthead{DEUTERIUM TOWARD $\delta$ ORI}
\begin{document}
\title{Spatial Variability in the Ratio of Interstellar Atomic Deuterium
to Hydrogen. I. Observations toward $\delta$ Orionis by the Interstellar
Medium Absorption Profile Spectrograph}
\author{Edward B. Jenkins, Todd M. Tripp, Przemys{\l}aw R. Wo\'zniak,}
\affil{Princeton University Observatory, Princeton, NJ 08544}
\author{Ulysses J. Sofia\altaffilmark{1}}
\affil{Department of Astronomy and Astrophysics, 
Villanova University, Villanova, PA 19085}
\and
\author{George Sonneborn}
\affil{Code 681, NASA Goddard Space Flight Center, Greenbelt, MD 20771}
\altaffiltext{1}{Present address: Dept. of Astronomy, Whitman College,
345 Boyer Ave., Walla Walla, WA 99362.}
\begin{abstract}
Studies of the abundances of deuterium in different astrophysical sites
are of fundamental importance to answering the question about how much
deuterium was produced during big bang nucleosynthesis and what fraction
of it was destroyed later.  With this in mind, we used the Interstellar
Medium Absorption Profile Spectrograph (IMAPS) on the ORFEUS-SPAS~II
mission to observe at a wavelength resolution of $4\,{\rm km~s}^{-1}$
(FWHM) the L$\delta$ and L$\epsilon$ absorption features produced by
interstellar atomic deuterium in the spectrum of $\delta$~Ori~A.  A
$\chi^2$ analysis indicated that $0.96 < N({\rm D~I})< 1.45\times
10^{15}\,{\rm cm}^{-2}$ at a 90\% level of confidence, and the gas is at
a temperature of about 6000K.  In deriving these results, we created a
template for the velocity profile defined by 7 different N~I transitions
recorded at a high signal-to-noise ratio.  Extra free parameters in the
analysis allowed for the additional uncertainties that could arise from
various sources of systematic error.

To derive a value for D/H, we measured the L$\alpha$ absorption features
in 57 spectra of $\delta$~Ori in the IUE archive, with the objective of
arriving at a more accurate H~I column density than those reported by
other investigators.  From our measurement of $N({\rm H~I})= 1.56\times
10^{20}\,{\rm cm}^{-2}$, we found that $N({\rm D~I})/N({\rm H~I})=
7.4^{+1.9}_{-1.3}\times 10^{-6}$ (90\% confidence).  Systematic errors
in the derivation of $N$(H~I) probably dominate over the very small
formal error, but their relative value should be smaller than that for
$N$(D~I).  Our result for D/H contrasts with the more general finding
along other lines of sight that ${\rm D/H} \approx 1.5\times 10^{-5}$. 
The underabundance of D toward $\delta$~Ori~A is not accompanied by an
overabundance of N or O relative to H, as one might expect if the gas
were subjected to more stellar processing than usual.
\end{abstract}


\section{Introduction}\label{introduction}

The relative abundances of the light elements not only substantiate the
standard interpretation for Big Bang Nucleosynthesis\footnote{However
see Gnedin \& Ostriker \protect\markcite{2347} (1992) and Burbidge \&
Hoyle \markcite{3646} (1998) for contemporary
viewpoints that differ from this interpretation.} (BBN) \markcite{3601,
3599} (Reeves et al. 1973; Epstein, Lattimer, \& Schramm 1976), but they
also hold the key for our determining the universal ratio of baryons to
photons, commonly designated by the parameter $\eta$ \markcite{3521,
3600, 3602} (Boesgaard \& Steigman 1985; Olive et al. 1990; Smith,
Kawano, \& Malaney 1993). There has been considerable interest in
measuring the abundance of deuterium, since its production was strongly
regulated by photodestruction in the radiation bath during the BBN,
making D/H a strong discriminant of $\eta$.

Deuterium is also destroyed in stars.  After having passed through one
or more generations of stars, diffuse gases that we can observe have
probably had their deuterium abundances reduced to values below those
that result from BBN.  Thus it is important to observe systems that have
different levels of chemical enrichment and mixing\markcite{4101}
(Timmes et al. 1997), so that we can untangle the effects of the two
fundamental destruction mechanisms, i.e., the photodestruction
accompanying BBN and the astration of material as the universe matures. 
A key step in this area of research is to form a solid foundation of
measurements of D in the chemically evolved gas in the disk of our
Galaxy.   Ultimately, when these results are combined with
determinations for distant gas systems that have not aged as much, we
expect to achieve a better understanding about the processing of gas
through stars, which is interesting in its own right \markcite{2366,
3098, 3308, 3611, 3604} (Steigman \& Tosi 1992, 1995; Dearborn,
Steigman, \& Tosi 1996; Scully et al. 1997; Tosi et al. 1998), and this
in turn should allow us to extrapolate the concentration of D back to an
era very soon after its primordial production.

An important foundation in recognizing the relationship between D/H and
some measure of stellar processing, such as the relative abundances of
elements produced in stellar interiors, is that an empirical
relationship between the two forms a unique sequence.  If this turns out
not to be true, then more elaborate interpretations of chemical
evolution may be needed.  To explore this issue, we have embarked on a
program to revisit some lines of sight studied by other investigators
(\S\ref{bkgnd}), but this time using much higher resolution spectra
obtained with the Interstellar Medium Absorption Profile Spectrograph
(IMAPS).

In this paper, we investigate the spectrum of $\delta$~Ori~A (HD 36486). 
This star has a spectral classification of O9.5~II, is a spectroscopic
binary, and is a member of the Ori OB1 association that has a distance
modulus of 8.5 ($d=500\,{\rm pc}$) \markcite{3657} (Humphreys 1978).  In
a companion paper \markcite{S99}(Sonneborn et al. 1999) we will report
on results for $\gamma^2$~Vel and $\zeta$~Pup.  The basic properties of
our spectrum of $\delta$~Ori and the instrument that recorded it are
discussed in \S\ref{basic}, followed by examinations of systematic
errors that could arise in the determination of a very weak
contamination signal (\S\ref{contam}), the intensity of the scattered
light background (\S\ref{bkg}), and absorption features from other
species (\S\ref{interference}).  In \S\ref{template} we describe how we
obtained independent information on the shape of the velocity profile
for material toward $\delta$~Ori, so that we could undertake our study
with only a small number of unknown, free parameters.  We have paid
considerable attention to minimizing the errors and evaluating them in a
fair and consistent manner (\S\ref{errors}).  For our value of $N$(D~I)
reported in \S\ref{N(DI)} to be useful, we must compare it with
$N$(H~I), and we must strive to make the accuracy of the latter as good
as or better than the former.  In \S\ref{N(HI)} we discuss our
comprehensive investigation of the IUE archival data that show L$\alpha$
absorption in the spectrum of $\delta$~Ori.  This special analysis
combined with our determination of $N$(D~I) ultimately led to our
determination of the atomic D/H toward $\delta$~Ori reported in
\S\ref{D/H}.  We relate this result to the abundances of other elements
relative to H in \S\ref{heavy} and discuss its significance in
\S\ref{disc}. 

\section{Previous Measurements of Atomic D/H in the Galaxy}\label{bkgnd}

Early measurements of D/H obtained from the {\it Copernicus\/} satellite
(resolution $15\,{\rm km~s}^{-1}$ FWHM) and IUE (resolution $25\,{\rm
km~s}^{-1}$ FWHM), summarized by Vidal-Madjar \& Gry \markcite{3510}
(1984), had a few cases that differed by more than the reported errors
from a general average ${\rm D/H}\approx 1.5\times 10^{-5}$.  At face
value, this suggested that D/H varies from one location to the next. 
McCullough \markcite{178} (1992) revisited this problem and asserted
that the evidence for such variations was not convincing.  In making his
claim that all of the data were consistent with a constant value for
D/H, McCullough rejected all of the deviant cases on the grounds that
the complexity of their velocity structures made the measurements much
less accurate than originally claimed.

The high resolution ($2.5\,{\rm km~s}^{-1}$ FWHM) and good sensitivity
of the GHRS instrument on HST enabled an accumulation of very accurate
observations of the interstellar L$\alpha$ H and D absorption features
superposed on the broader chromospheric L$\alpha$ emission lines of
nearby F, G and K type stars.  The best determinations of D/H were those
toward $\alpha$~Aur (Capella), where Linsky, et al. \markcite{3085}
(1995) found that ${\rm D/H}=1.60_{-0.19}^{+0.14}\times 10^{-5}$, and
toward HR 1099, where Piskunov, et al. \markcite{4220} (1997) obtained
${\rm D/H}=1.46\pm 0.09\times 10^{-5}$.  The issue of whether or not
atomic deuterium to hydrogen ratios toward other cool stars differ from
these values has been an elusive one, although it seems clear that one
could rule out deviations greater than about 50\% in either direction
\markcite{3119, 4220, 3467} (Wood, Alexander, \& Linsky 1996; Dring et
al. 1997; Piskunov et al. 1997).  The chief problem has been that the
measurements of $N({\rm H~I})$ toward late-type stars were very
dependent on assumptions about the shape of the underlying emission
profile \markcite{4220, 3118} (Linsky \& Wood 1996; Piskunov et al.
1997) or the compensations for additional, broad absorptions caused by
hydrogen walls associated with the stellar wind cavities around either
the Sun or the target stars \markcite{3118, 4279} (Linsky \& Wood 1996;
Wood \& Linsky 1998).  Even so, these investigations revealed some
intriguing, convincing variations for the abundances of D~I with respect
to those of Mg~II.  Unfortunately, the significance of these changes is
clouded by the possibility that they could result simply from
alterations in the amount of depletion of Mg onto dust grains
\markcite{1053, 1063, 2722 4210} (Murray et al. 1984; Jenkins, Savage,
\& Spitzer 1986; Sofia, Cardelli, \& Savage 1994; Fitzpatrick 1997).

Lemoine, et al \markcite{3258} (1996) observed the interstellar H and D
L$\alpha$ absorption features in the spectrum of the DA white dwarf
G191$-$B2B with the GHRS and reported their determinations for D/H. 
Later, high-resolution observations by Vidal-Madjar, et al
\markcite{3562} (1998) brought forth some refinements in the
interpretation of the velocity structures of the absorption profiles,
leading to a determination ${\rm D/H} = 1.12\pm 0.08\times 10^{-5}$ for
all of the material in front of this star.  If one allows for the fact
that a contribution from the Local Interstellar Cloud (LIC) is somewhat
blended with those of more distant clouds and adopts the $\alpha$~Aur
result for the LIC, D/H toward the other material could be of order
$9\times 10^{-6}$.  This low value for D/H is supported by observations
of the hot subdwarf BD~$+28\arcdeg 4211$ reported by G\"olz, et al.
\markcite{3577} (1998), ${\rm D/H}=8_{-4}^{+7}\times 10^{-6}$, although
the error bar is large enough to include the results obtained for
$\alpha$~Aur, HR~1099, and other late-type stars.

For lines of sight that have hydrogen column densities that are small
enough to analyze using the L$\alpha$ profile, there is the danger that
improper allowances for either L$\alpha$ emission (cool stars) or
absorption (hot dwarfs) could lead to errors.  Moreover, in some
circumstances hydrogen walls associated with either the target stars or
the Sun can lead to complications.  One way to bypass these problems is
to examine the higher Lyman series absorption features toward more
distant, early-type stars with much more foreground material, as was
done with the {\it Copernicus\/} satellite.  We also have the benefit of
sampling the interstellar medium well outside our immediate vicinity. 
However, a principal weakness of {\it Copernicus\/} was its limited
resolving power ($\approx 15~{\rm km~s}^{-1}$ FWHM).\footnote{Another
drawback with {\it Copernicus\/} was that all observations had to be
taken in sequence, since the spectrometer was a scanning instrument. 
Vidal-Madjar, et al. \protect\markcite{3509} (1982) obtained
inconsistent results for different Lyman series lines in the spectrum of
$\epsilon$~Per, an effect which they attributed to the influence of
stellar features that varied with time.}  In large part, the {\it
Copernicus\/} investigators had to model the instrumentally smeared,
detailed velocity structure of the gas, with guidance from
high-resolution observations of Na~I features recorded from the ground. 
Unfortunately, the sodium D lines are a poor standard for comparison
because their strengths are dependent on ionization equilibria that are
entirely different from those of D~I and H~I.  In this study we revisit
the case for $\delta$~Ori, originally observed with {\it Copernicus\/}
by Laurent, et al. \markcite{1784} (1979), but now with new observations
taken with an instrument with considerably better velocity resolution
than {\it Copernicus}.

\section{Observations and Data Reduction}\label{obs}

\subsection{Basic Properties of the Spectra}\label{basic}

A far-UV spectrum of $\delta$~Ori over the wavelength interval 930 to
1150\,\AA\ was recorded in a series of exposures lasting 54 min over
various observing intervals between 22 November and 3 December 1996 by
the Interstellar Medium Absorption Profile Spectrograph (IMAPS).  This
series of observations was undertaken during the ORFEUS-SPAS~II mission
\markcite{3532} (Hurwitz et al. 1998) on STS-80, which was the second
orbital flight of IMAPS. IMAPS is a simple, objective-grating echelle
spectrograph that can record the far-UV spectrum of a bright, early-type
star with sufficient resolution to show many of the velocity structures
in the interstellar lines.  Jenkins, et al. \markcite{340} (1996)
present a detailed description of the IMAPS instrument, its performance
on the first ORFEUS-SPAS mission in 1993\footnote{Improvements in IMAPS
after the first flight removed most of the problems discussed by
Jenkins, et al. \protect\markcite{340}(1996; in particular, see their
\S8.2).  Most important, the severe changes in photocathode sensitivity
that were evident on the first flight were not manifested on the second
flight.}, and the methods of data correction and analysis.

We summarize very briefly how the spectra are recorded by IMAPS: In any
single exposure that covers an angle $18\arcmin 20\arcsec \times
14\arcmin 40\arcsec$, one-quarter of the echelle grating's free spectral
range and, nominally, diffraction orders 194 through 242 are recorded by
an electron-bombarded CCD image sensor.  This detector has an opaque
photocathode on a smooth substrate and uses magnetic focusing to form
the electron image on the CCD.  Electrons impacting on the back side of
the specially thinned CCD have an energy of 18~keV, and they produce
enough secondary electrons within the silicon layer to make individual
photoevents appear as bright spots.  Each spot has an amplitude that is
about 20 times greater than the combined noise from the readout
amplifier and the random fluctuations in dark current. The CCD has a
format of $320\times 256$ pixels, each of which is $30 \mu$m square and
subtends a $\Delta\lambda$ equivalent to a Doppler shift of $1.25\,{\rm
km~s}^{-1}$.  The echelle orders are separated by about 5 CCD pixels,
but they are rather broad in the cross-dispersion direction.  The CCD is
read out 15 times a second.  The video signals from successive frames
are summed in an accumulating memory to produce the integrated spectral
images.  In our processing of these images after the flight, we
subtracted dark-current comparison frames that were recorded at frequent
intervals with the accelerating high voltage turned off.

\begin{figure}
\plotone{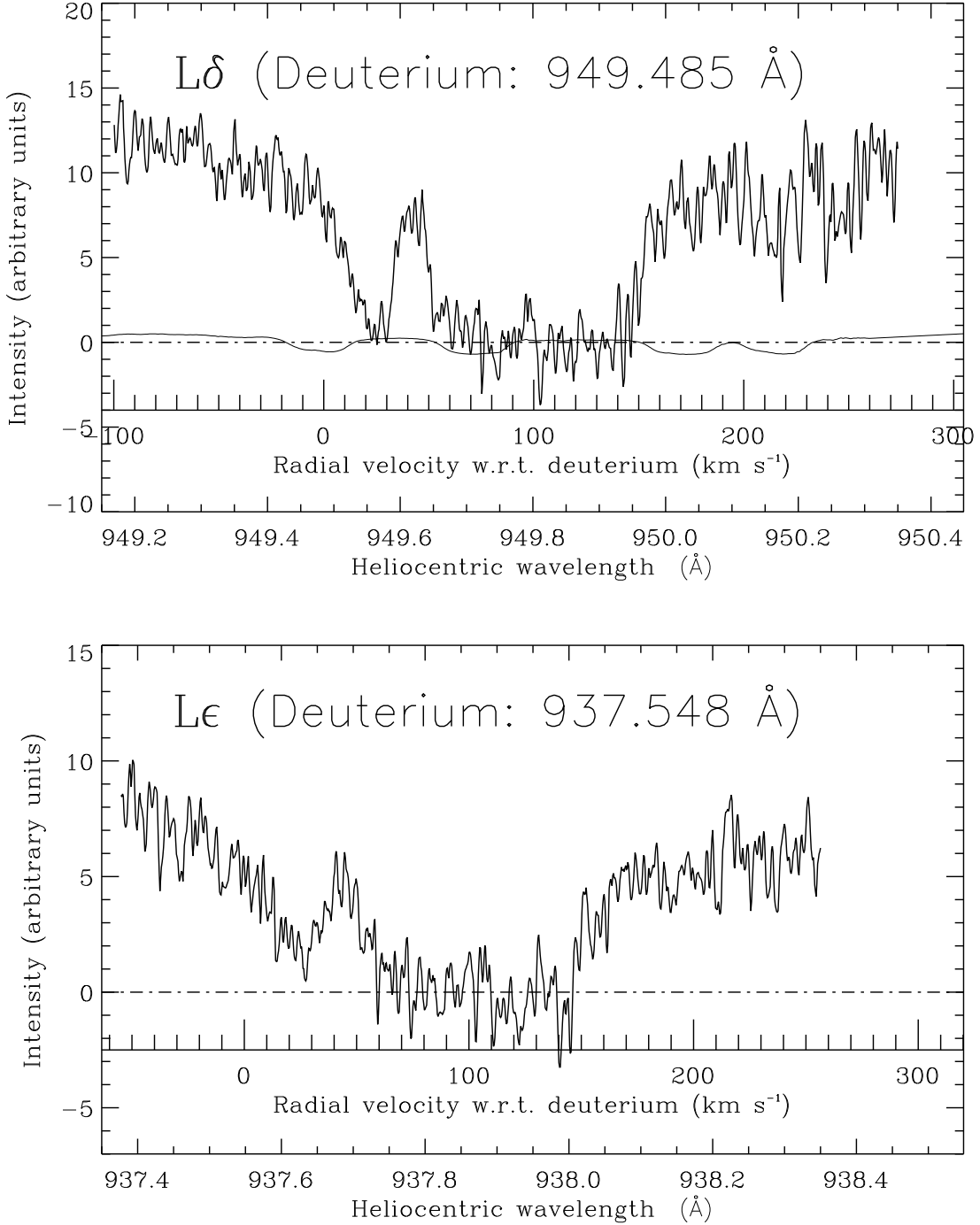}
\caption{The D and H absorption features at L$\delta$ and L$\epsilon$ in
the IMAPS spectrum of $\delta$~Ori, shown with a sampling interval of
\onehalf\ CCD pixel (\S\ref{basic}).  In both cases, the velocity scale
refers to the D feature in the heliocentric reference frame.  For
L$\delta$, the character and amplitude of a correction
(\S\protect\ref{contam}) for possible contamination from features in an
adjacent order is shown (line that oscillates on either side of the zero
level), with a scale factor $R_c=0.0070$ (best fits came out over the
range 0.0058 to 0.0080 for the largest extremes shown in
Table~\protect\ref{chi2_results}).  The spectra are also corrected for
the small effects arising from interfering lines
(\S\protect\ref{interference}).\label{dorideut}}
\end{figure}

The effective area of IMAPS on the 1996 flight was about $3\,{\rm cm}^2$
at wavelengths longward of about 1020\,\AA, leading to typical
signal-to-noise ratios of about 80 near the maximum of the echelle
grating's blaze angle for stars as bright as $\delta$~Ori~A.  However
the Al+LiF coatings on the two gratings have a low reflection efficiency
at shorter wavelengths, resulting in a factor of 10 lower effective area
in the vicinity of the L$\delta$ and L$\epsilon$ lines.  This reduced
efficiency coupled with the much lower flux at the centers of the
stellar Lyman series lines made it especially difficult to achieve high
values of signal-to-noise.  We overcame this problem by recording a
large number of spectra that could be added together.  The total
integrated flux at the continuum levels near L$\delta$ and L$\epsilon$
amounted to about 600 photons for each CCD pixel width in the dispersion
direction ($1.25\,{\rm km~s}^{-1}$).  Noise fluctuations in the spectra
had $rms$ deviations about equal to 1/10 of the continuum levels, with
the principal noise source being the multiple readouts of the CCD,
rather than from photon-counting statistical errors.  (This is clearly
evident in Fig.~\ref{dorideut}, which shows a noise level at zero
intensity to be about the same as that at the elevated intensity
levels.)

We deliberately introduced offsets in position for the spectra in
different sets of exposures.  This was done to reduce the possibility
that the spectrum could be perturbed by subtle flaws, such as CCD
columns with anomalous responses or variations in photocathode
efficiency with position (we could see no evidence for the latter
however).

\placefigure{dorideut}

Figure~\ref{dorideut} shows the D and H absorption profiles for
$\delta$~Ori at L$\delta$ and L$\epsilon$.  Observations of telluric
atomic oxygen lines from excited fine-structure levels, seen elsewhere
in the IMAPS spectrum of this star, indicated that the instrumental
profile that governs the wavelength resolution of these observations was
consistent with a Gaussian distribution having a FWHM\footnote{See
Jenkins \& Peimbert \protect\markcite{362} (1997) for the details on how
to arrive at this finding -- their measurements for the IMAPS spectrum
of $\zeta$~Ori~A are not far from those that apply to our $\delta$~Ori
spectrum.} equal to $4.0\,{\rm km~s}^{-1}$. At this resolving power, the
deuterium features are well separated from their hydrogen counterparts,
as can be seen in Fig.~\ref{dorideut}.  While the deuterium L$\delta$
profile does show some asymmetry, to within the uncertainties of the
noise there does not seem to be any extraordinary complexity in the
velocity structure of the D profiles.  For instance, there seems to be
no evidence for any strong, narrow velocity components buried within the
main peak of the L$\epsilon$ feature.  Most important, the D features
are not badly saturated.

\subsection{Possible Contamination Signal}\label{contam}

There is some overlap of signal from one echelle diffraction order to
the next.  Our optimal extraction routine was designed to compensate for
this effect \markcite{340} (Jenkins et al. 1996), but if this correction
was not perfect we may have had some contamination by the spectral
intensities from an adjacent order.  We had to be especially watchful
for this possibility in the vicinity of the Lyman series lines because
the stellar continuum level in the region of interest is much lower than
elsewhere.  This gives the contamination signal an advantage over the
signal we wished to study.  For L$\epsilon$ there is no problem because
there are no spectral features in the orders either directly above or
below the D and H features or their nearby continua.  Errors in
correcting for order overlap will only change the effective zero
intensity level, which is corrected out anyway.  The next higher order
of diffraction that appears just below the one that contains the
L$\delta$ absorptions is featureless, but, unfortunately, the order
above the L$\delta$ order exhibits interstellar absorption features from
the very strong multiplet of N~I at 954\,\AA.

In our investigation of the spectrum in the vicinity of L$\delta$, we
allowed for the possibility that our correction for the order overlap
was either too large or too small.  This error could have added a
spurious signal to our spectrum.  Therefore, we included as a free
parameter a scaling coefficient $R_c$ (which could be either positive or
negative) for the amplitude of a correction signal (with the same shape)
to cancel the possible residual contamination, and we allowed it to vary
as we explored for minimum values of $\chi^2$ (\S\S~\ref{errors} and
\ref{N(DI)}).  When this coefficient is less than 0.005 or larger than
0.03, unreasonably large perturbations can be seen in the bottom of the
very broad hydrogen feature.  Within this range, however, we allowed for
the fact that our derivation of $N$(D~I) could be influenced by the
exact value.   Figure~\ref{dorideut} shows the correction signal with
the most plausible amplitude, as indicated by minimum $\chi^2$ for $R_c$
at the most probable $N$(D~I) given in Table~\ref{chi2_results}.  The
spectrum that is shown in this figure has had this correction included.

\subsection{Background Level}\label{bkg}

Our analysis of the D profiles is very dependent on our having an
accurate determination of the level of zero intensity.  Sources of
background illumination include not only grating scatter, but also a
diffuse glow caused by a portion of the L$\alpha$ geocoronal background
that is not fully rejected by a mechanical collimator at the
instrument's entrance aperture.  (The detector's dark count rate is
negligible compared to these sources of background.)  Fortunately, we
could use the bottoms of the broad, heavily saturated H absorptions that
accompany each D profile establish the position of the background level.

In principle, the saturated portion of the H profile could mislead us if
there were broad, shallow wings in the instrumental profile caused by
scattering from the echelle grating.  If this were the case, one could
imagine that the local background level might increase slightly for
wavelengths somewhat removed from the strong H feature.  We can rule out
this prospect on two grounds.  First, before IMAPS was flown, we
illuminated it in a vacuum tank with a collimated beam from a molecular
hydrogen emission line source, and faint, very broad wings of the
recorded emission lines could be seen only on the strongest features. 
The energy in these wings corresponded to 15\% of the total, spread over
several \AA.  The remaining 85\% was within the main peak.  Second, for
$\delta$~Ori we found that for both L$\delta$ and L$\epsilon$ the
apparent depths of the D features in {\it Copernicus\/} scans (taken
with an ordinary grating in first order) showed excellent agreement with
those registered in the IMAPS spectrum after it had been degraded by
convolving it with the {\it Copernicus\/} instrumental profile function
[a triangle with FWHM = 0.045\,\AA\ \markcite{1784} (Laurent,
Vidal-Madjar, \& York 1979)].  For these two reasons, we feel confident that
the apparent flux in the bottom of the H feature is, to within the
uncertainties from noise, a good representation for the zero level under
the deuterium line.

\subsection{Interference from Other Lines}\label{interference}

Table~\ref{other_lines} lists lines from species other than D~I and H~I
that are in vicinity of the deuterium absorptions or the fragments of
the spectrum that were used to define the continuum level
(\S\ref{N(DI)}).  The Werner 3$-$0~P(2) line lies within the H L$\delta$
feature, and thus it is of no importance.   From other lines out of the
$J$=4 level of H$_2$ that appear elsewhere in our IMAPS spectrum, we
know that the 4$-$0~P(4) line of the Werner system should have a
negligible strength.  Again using other features in the IMAPS spectrum,
we found that the remaining two lines shown in the table could perturb
our spectra and influence our final results.  We therefore felt it was
necessary to estimate their strengths and then apply a correction to
compensate for their presence.

\placetable{other_lines}
\begin{deluxetable}{lccl}
\tablecaption{Lines in the Vicinity of Deuterium L$\delta$ and
L$\epsilon$\label{other_lines}}
\tablewidth{0pt}
\tablehead{
\colhead{Identification} & \colhead{$\lambda$\tablenotemark{a}} &
\colhead{$\log f\lambda$\tablenotemark{b}} & \colhead{Comment}\\
\colhead{} & \colhead{(\AA)} & \colhead{} & \colhead{}}
\startdata
H$_2$ Werner 3$-$0 P(2)&949.608&$-0.078$&Within the L$\delta$ H~I
absorption\nl
H$_2$ Lyman 14$-$0 P(2)&949.351&0.95&On cont. left of the L$\delta$ D~I
absorption\nl
Fe~II\dotfill&937.652&1.263&On cont. between L$\epsilon$ D~I and H~I
absorptions\nl
H$_2$ Werner 4$-$0 P(4)&937.551&0.89&Too weak to matter\nl
\enddata
\tablenotetext{a}{The H$_2$ Lyman lines are from Abgrall et al.
\protect\markcite{280} (1993a), the Werner lines are from Abgrall et al.
\protect\markcite{281} (1993b) and the Fe~II line is from Morton
\protect\markcite{96} (1991).  The maximum absorptions should occur at a
Doppler shift of about +23 to $+25\,{\rm km~s}^{-1}$ with respect to the
heliocentric wavelength scales shown in Fig.~\protect\ref{dorideut}.}
\tablenotetext{b}{All of the H$_2$ lines are from Abgrall \& Roueff
\protect\markcite{2066} (1989) and the Fe~II line is from Morton
\protect\markcite{96} (1991).}
\end{deluxetable}
\clearpage

To estimate the strength of the 14$-$0~P(2) Lyman line of H$_2$, we
noted that the 4$-$0~R(2) line at 1051.498\,\AA\ had a maximum depth of
0.27 times the local continuum at $v=23\,{\rm km~s}^{-1}$, and it was
recorded in a part of our spectrum where the signal-to-noise ratio was
about 80.  This line has a value for $f\lambda$ that is 1.7 times that
of the 14$-$0~P(2) line.  To compensate for the effect of the latter on
our continuum to the left of the D~I L$\delta$ feature, we divided the
observed spectrum by the continuum-normalized intensities in Lyman
4$-$0~R(2) profile all taken to the 1/1.7 power, with the profile
shifted in wavelength to match that of the 14$-$0 P(2) line.

For a template of the Fe~II absorption, we used the line at
1081.875\,\AA\ that shows a maximum depth of 0.26 at $v=25\,{\rm
km~s}^{-1}$ recorded at S/N~=~90 (this maximum for the 937.652\,\AA\
line falls within the H absorption) and a shoulder at $v=12\,{\rm
km~s}^{-1}$ with a depth of 0.12.  This shoulder for the 937.652\,\ line
falls on top of a critical piece of continuum between the D~I and H~I
L$\epsilon$ features.  The 1081.875\,\ line has $f\lambda$ that is 0.82
times that of the interfering feature \markcite{96} (Morton 1991), and
once again this difference was taken into account when we made the
correction.  

\section{Interpretation of the Data}\label{interpretation}

\subsection{Velocity Profile Template}\label{template}

To derive the most accurate value for the column density of atomic
deuterium $N$(D~I), it is beneficial to use information from other
species recorded at much higher S/N to help define more accurately the
shape of the D~I velocity profile.  Profiles of O~I and N~I are probably
the most suitable comparison examples for two reasons.  First, these two
elements have very mild depletions, if any, caused by the atoms
condensing into solid form onto dust grains \markcite{3466, 4196}
(Meyer, Cardelli, \& Sofia 1997; Meyer, Jura, \& Cardelli 1998).  The
column densities of N and O seem to track those of H over a diverse
sample of regions \markcite{1038, 1073} (Ferlet 1981; York et al. 1983). 
As a consequence, it is unlikely that their velocity profiles will
differ appreciably from that of D~I.  This is in contrast to the usual
striking differences exhibited between elements that are mildly
depleted, such as Na~I, and elements that are strongly depleted, such as
Ca~II. The former is generally concentrated at lower velocities than the
latter for a given line of sight \markcite{2391, 2392, 2481, 2754}
(Routly \& Spitzer 1952; Siluk \& Silk 1974; Vallerga et al. 1993;
Sembach \& Danks 1994).  Second, O~I and N~I have ionization potentials
close to that of neutral hydrogen, and this close match in energy makes
their susceptibility to ionization nearly the same and also insures that
the cross sections for (nearly resonant) charge exchange are high
\markcite{3406, 1927} (Field \& Steigman 1971; Butler \& Dalgarno 1979). 
For this reason, plus the consideration that whatever means there are
for ionizing N and O will operate in much the same way for H (or D), we
can generally regard the relative ionizations of oxygen and nitrogen to
be good representations for that of D [but for evidence to the contrary,
see Vidal-Madjar et al.   \markcite{3562} (1998)].  This assumes, of
course, that O and N are not being ionized appreciably to multiply
charged states.\footnote{We looked for absorption by the N~III
transition at 989.799\,\AA\ in our IMAPS spectrum of $\delta$~Ori~A.  No
absorption was evident at $v=25\,{\rm km~s}^{-1}$, but it was difficult
to assign a quantitative upper limit because of interference from the
nearby feature of Si~II at 989.873\,\AA.}

In the wavelength coverage of IMAPS where we have a reasonably good S/N,
there are no transitions from the O~I ground state that are weak enough
(and with known f-values) to yield absorption lines that we can analyze. 
There is, however, a good series of exposures in the HST
archive\footnote{Exposure identifications z2zb0304t, z2zb0305t and
z2zb0306t.} that cover the O~I 1355.6\,\AA\ feature in the spectrum of
$\delta$~Ori, recorded by the GHRS Echelle-A spectrograph. 
Unfortunately, the transition probability for this line is so weak that
only the main peak in the velocity profile shows up above the noise.  

For N~I, within the coverage of IMAPS there are three multiplets (at
952.4\,\AA, 953.8\,\AA\ and 954.1\,\AA) from the ground level that are
ideal for studying the apparent distribution\footnote{For the
distinction between the {\it apparent\/} and {\it true\/} velocity
distributions, see the discussions by Savage \& Sembach
\protect\markcite{110} (1991) and Jenkins \protect\markcite{3184}
(1996).  In our study of $\delta$~Ori at high velocity resolution, it is
probably safe to assume that the two are equal to each other.} $N_a(v)$
of the nitrogen atoms with velocity, defined as
\begin{equation}\label{N_a}
N_a(v) = 3.768\times 10^{14}{\tau_a(v)\over f\lambda}{\rm cm}^{-2}({\rm
km~s}^{-1})^{-1}~,
\end{equation}
where the apparent optical depth $\tau_a(v)$ is a valid quantity to
measure at velocities $v$ where the line is not badly saturated or,
alternatively, not too weak.  In this equation $f$ is the transition's
oscillator strength, and $\lambda$ is expressed in \AA.  In our study of
the N~I lines, we adopted f-values from the laboratory measurements of
Goldbach, et al. \markcite{3555} (1992). For the triplet at 952.4\,\AA,
the weakest feature at 952.523\,\AA\ is only moderately saturated.  The
other two features are heavily saturated but useful for revealing the
weaker shoulder on the left-hand side of the main peak.  The 4 much
stronger features of N~I in the vicinity of 954\,\AA\ are useful for
defining accurately the behavior of $N_a(v)$ at velocities where it is
below about $5\times 10^{13}{\rm cm}^{-2}({\rm km~s}^{-1})^{-1}$.  We
derived a composite $N_a(v)$ profile for N~I from the 7 lines using the
method employed by Jenkins \& Peimbert \markcite{362} (1997) when they
synthesized the profiles of H$_2$ in various $J$ levels toward
$\zeta$~Ori~A.

\begin{figure}
\plotone{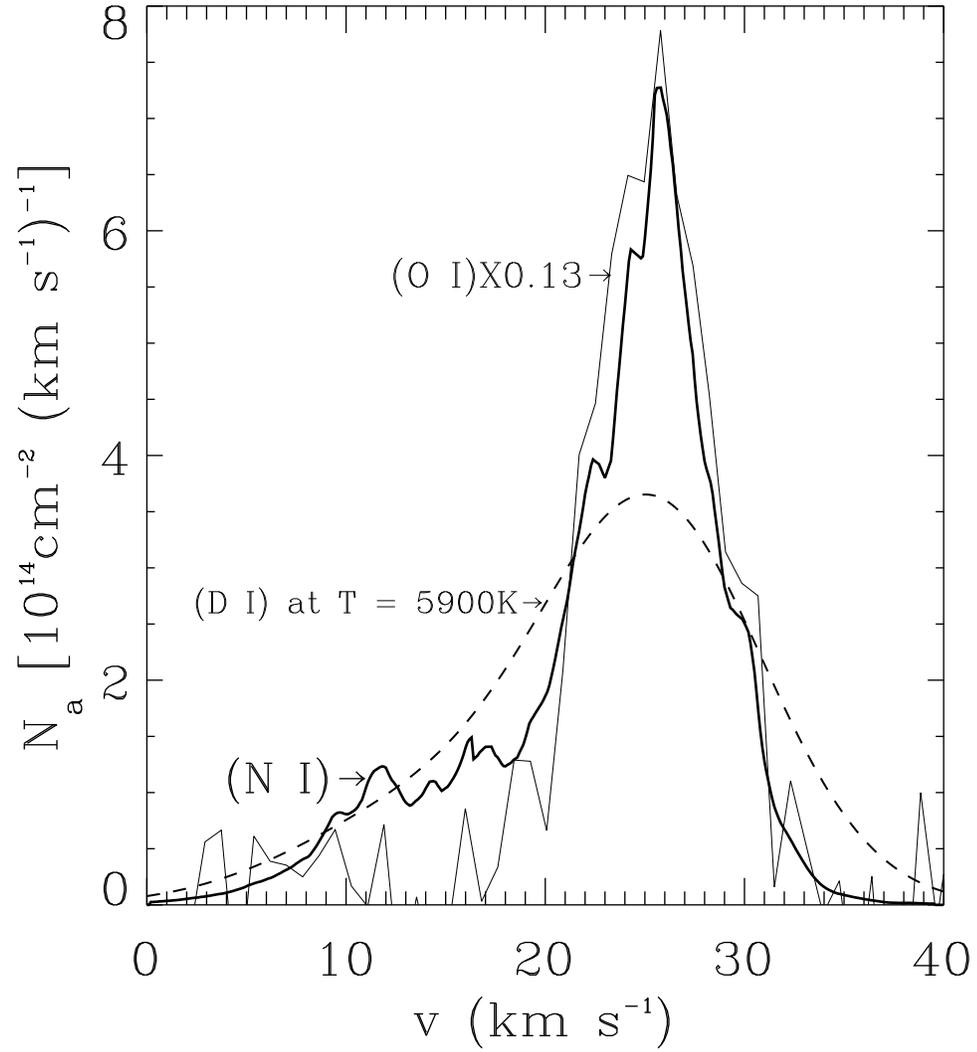}
\caption{Profiles for $N_a(v)$ for N~I (heavy line) and O~I (light
line), recorded by IMAPS and HST, respectively.  The dashed line shows
the expected shape of the deuterium profile for an additional thermal
broadening at $T=5900$K (see Eq.~\protect\ref{convol}) of the N~I
profile favored by our most likely solution in
Table~\protect\ref{chi2_results}.\label{NandO}}
\end{figure}

Figure~\ref{NandO} shows the $N_a$ profiles that we derived for N~I and
O~I.  We chose to work with the N~I profile in our interpretation of the
D lines because it was of much better quality.  This pragmatic reason
for choosing N~I as a template is contrary to the idealistic stance that
O~I would be a better match to D~I, based on evidence from the
absorption lines in the spectrum of G191-B2B recorded by Vidal-Madjar,
et al \markcite{3562} (1998) and calculations of partially ionized
atomic gases by Sofia \& Jenkins \markcite{4310} (1998).  We note,
however, that to within the noise fluctuations the significant part of
the profile of O~I is consistent with the main part of the profile of
N~I.  There is a suggestion of an apparent inconsistency between N~I and
O~I in the velocity range $10 < v < 20\,{\rm km~s}^{-1}$.  While this
may be true, we point out that this weak shoulder in the O~I absorption,
if it exists, may have been lost in the fitting of the continuum to the
curvature of the stellar spectrum, which is larger than the expected
size of the shoulder indicated by the N~I profile.   Also, the existence
of some absorption to the left of a main peak is supported by the shape
of the L$\delta$ deuterium profile shown in Figures \ref{dorideut} and
\ref{extremes}.

\placefigure{NandO}

In principle, we should be cautious about possible contamination of the
interstellar N~I profile by nitrogen atoms in the Earth's atmosphere
above our orbital altitude of 295\,km.  Above this altitude, the MSIS-86
model atmosphere for solar minimum shows an exponential decrease in the
density of nitrogen atoms with a scale height of 53\,km, starting at a
density of $5.5\times 10^6{\rm cm}^{-3}$ at 295\,km \markcite{2799}
(Meier 1991).  For a zenith angle $z$ of 45\arcdeg, we calculate that
the telluric contribution to the observed $N$(N~I) should be $1.3\times
10^{13}{\rm cm}^{-2}$, an amount that would be just about invisible in
the representation shown in Fig.~\ref{NandO}.  Even at $z=90\arcdeg$,
$N({\rm N~I})=1.3\times 10^{14}{\rm cm}^{-2}$, which is just slightly
larger than the bump (presumably due to noise) immediately to the right
of the ``(N~I)$\rightarrow$'' indication in the figure.  Thus, we feel
it is safe to dismiss the possibility that any telluric N~I
contamination is large enough to influence our profile.

One important difference between D and either N or O is the atomic mass. 
If the thermal Doppler broadening for deuterium atoms is not very much
less than that due to macroscopic motions, we would expect the D
profiles to be broader than those of O and N.  With the simplifying
assumption that the temperature of the gas does not vary much from place
to place, we expect that for nitrogen the observed distribution of the
atoms with velocity is represented by the turbulent motions $t(v)$
convolved with the thermal Doppler profile, $\phi_D(m,T,v)$, given by
\begin{equation}\label{d}
\phi_D(m,T,v)=\sqrt{m/(2\pi kT)}\exp[-mv^2/(2kT)]~,
\end{equation}
with $m$ equal to 14 times the proton mass $m_p$.  For convenience, we
can include in $t(v)$ the instrumental smearing of the profile, on the
condition that we are not being misled by saturated, unresolved
structures in the absorption line \markcite{110, 3184} (Savage \&
Sembach 1991; Jenkins 1996).  The same relation holds for deuterium with
$m=2m_p$.  Since the convolution of two Gaussian distributions produces
a third with a second moment equal to the sum of the two original ones,
we can state that
\begin{equation}\label{convol}
t(v)\ast \phi_D(2m_p,T,v)=t(v)\ast \phi_D(14m_p,T,v)\ast
\phi_D(7m_p/3,T,v)~.
\end{equation}

When we analyzed the deuterium features, we adopted for a standard model
of their shapes the nitrogen velocity profile convolved with the last
term in Eq.~\ref{convol}.  We allowed the temperature $T$ to be a free
parameter that could influence the fit between the profiles of N~I and
D~I (and one that is also of some astrophysical relevance).  We did not
allow for variations of $T$ among unrecognized and blended velocity
substructures that contributed to the profile, since the identification
of these components is somewhat arbitrary.  Our goal was to account for
the general modification of the profile due to the known differences in
the effects of thermal and turbulent broadening. (For determining
$N$(D~I), the weakness of the dependence of the derived $N$(D~I) with
$T$ expressed in the endnote of Table~\ref{chi2_results} indicates that
our simplification that $T$ is constant is probably safe.) 
Figure~\ref{NandO} shows our model for the deuterium profile (smooth,
dashed line) for a value of $T$ that gave the minimum $\chi^2$ for the
preferred value of $N$(D~I) in Table~\ref{chi2_results}.  This is a
smoothed version of the N~I profile (heavy, solid line) that was
obtained from the convolution by the kernel $\phi_D(7m_p/3,T,v)$ from
Eq.~\ref{convol}.

In addition to allowing $T$ to vary, we also allowed for the existence
of a uniform velocity offset between N~I and D~I, in recognition of the
possibility that either our wavelength scale or the laboratory
wavelengths of the N~I features had some small, systematic errors.

\subsection{Allowances for Random and Systematic Errors}\label{errors}

The presence of random errors due to noise fluctuations in the signal
presents the usual challenge of determining the most probable result for
$N$(D~I) and permissible variations that still give an acceptable fit to
the data.  On top of this we must consider additional uncertainties
caused by systematic errors.  The ones that we can identify easily are
the inaccuracies in defining the background level for both lines
(\S\ref{bkg}) and the contamination signal in the L$\delta$
profile(\S\ref{contam}).  Additional parameters that could affect the
outcome are the temperature of the gas $T$ through its effect in making
the deuterium profile smoother than the N~I template (\S\ref{template}),
the difference in the zero points of the N~I and D~I velocity scales,
and the adopted heights and slopes of the continuum levels over the
deuterium L$\delta$ and L$\epsilon$ absorption features.  Our tactic for
coping with these systematic errors was to express them in terms of
simple parameters that could vary and then consider them in a unified
analysis.  Since these errors could be correlated, we felt that it would
be unwise to analyze them separately.

We determined how well the data conform to various combinations of
parameter values by evaluating the conventional $\chi^2$ statistic,
$\sum [(I_{\rm meas}-I_{\rm exp})/\sigma_I]^2$, where $I_{\rm meas}$ is
a measured intensity with uncertainty $\sigma_I$, and $I_{\rm exp}$ is
the expected intensity given the specified set of parameter values.  
Useful discussions of how to interpret the determinations of $\chi^2$
when there are many free parameters are given by Lampton, et al.
\markcite{1666} (1976) and Bevington \& Robinson \markcite{3528} (1992,
p. 212).  The basic scheme is to find the minimum value $\chi^2({\rm
min})$ that arises when all parameters that influence $I_{\rm exp}$ are
free to vary, and then examine the deviations $\chi^2-\chi^2({\rm min})$
as the parameters stray from their optimum values.  Our $\chi^2$ values
represented a sum over both the L$\delta$ and L$\epsilon$ features taken
together.  This approach is similar to one adopted by Burles \& Tytler
\markcite{4365} (1998a) in their analysis of the deuterium abundance in
a quasar absorption line system, except that we decouple the hydrogen
measurements (\S\ref{N(HI)}) from those of deuterium because they are
fundamentally different from each other.

To limit the number of degrees of freedom that apply to the confidence
intervals for the outcomes that we are interested in, we segregated the
free parameters into two fundamental categories.  First, we recognized
those parameters that had an astrophysical significance, $N$(D~I) and
$T$.  We sought to find a confidence interval that constrained these two
parameters simultaneously.  While $T$ might seem to be an incidental
parameter outside the objective of this study, there were good physical
reasons for our verifying that no appreciable portion of the probability
density wandered above or below acceptable limits.  The second category
contained parameters that were of no particular interest to us, i.e.,
nuisance parameters, but ones that had to be allowed to change freely as
we re-minimized the $\chi^2$ for every new trial combination of $N$(D~I)
and $T$.  We had no profound reason to require that any of these
variables in the second category be constrained, and thus we could
consider a projection of the lowest $\chi^2$ values in the
multi-dimensional space of these variables onto just the $N$(D~I)$-T$
plane.  This allowed us to restrict the number of the degrees of freedom
(df) that applied to the confidence intervals down to only 2.

In summary, parameters that mattered were (1) $N$(D~I) and (2) $T$,
while those that did not were (3) the coefficient $R_c$ for scaling the
contamination signal (\S\ref{contam}), (4) a relative velocity error
$\Delta v_{\rm N,D}$ between the N~I profile template (\S\ref{template})
and the deuterium absorption, (5 and 6) the background levels in the
bottom of the H L$\delta$ and L$\epsilon$ features, and finally (7
through 10) the two coefficients that described the continuum straight
lines spanning the D L$\delta$ and L$\epsilon$ features, i.e., in each
case the level near the middle of the D feature and the slope of the
line.\footnote{While one might argue that the continuum is not straight
and has a curvature produced by the damping wings of the H~I absorption,
this effect is probably small enough to be masked by a curvature in the
opposite direction caused by the broad stellar hydrogen feature.  For
the value of $N$(H~I) given in \S\protect\ref{N(HI)}, the damping wing
absorbs only 9\% of the flux at the D L$\delta$ line and 2\% at the D
L$\epsilon$ line.  Moreover, the {\it appearance} of the continuum
suggests that a straight-line fit is justified -- see
Fig.~\protect\ref{extremes}.}

\subsection{Determination of $N$(D~I) and its Uncertainty}\label{N(DI)}

From our knowledge of the CCD readout noise and dark current combined
with the statistical fluctuations in the (background + signal) photons,
we made an initial estimate for the uncertainties in the individual
measurements of intensity $\sigma_I$ at each velocity.  We determined
the correlation length for these errors by comparing fluctuations of
intensity at a velocity $v$ with those of $v+\Delta v$.  The
correlations disappear for $\Delta v=1.25\,{\rm km~s}^{-1}$, which is
exactly the width of each pixel in the CCD.\footnote{This is not a
trivial finding.  In an electron-bombarded CCD image sensor, correlation
lengths greater than a CCD pixel can result if the diameter of the
secondary electron cloud from each event is of order or larger than a
CCD pixel \protect\markcite{1390} (Jenkins et al. 1988).  Events
straddling a pixel boundary, for instance, will create a correlated
signal in both pixels that pick up the secondary charges.  Evidently
such events are not common enough to cause a statistically significant
effect, otherwise we would have found correlation lengths greater than
$1.25\,{\rm km~s}^{-1}$.}  Our determinations of $\chi^2$ discussed
below relied on intensities separated by this value for $\Delta v$.

\begin{figure}
\plotone{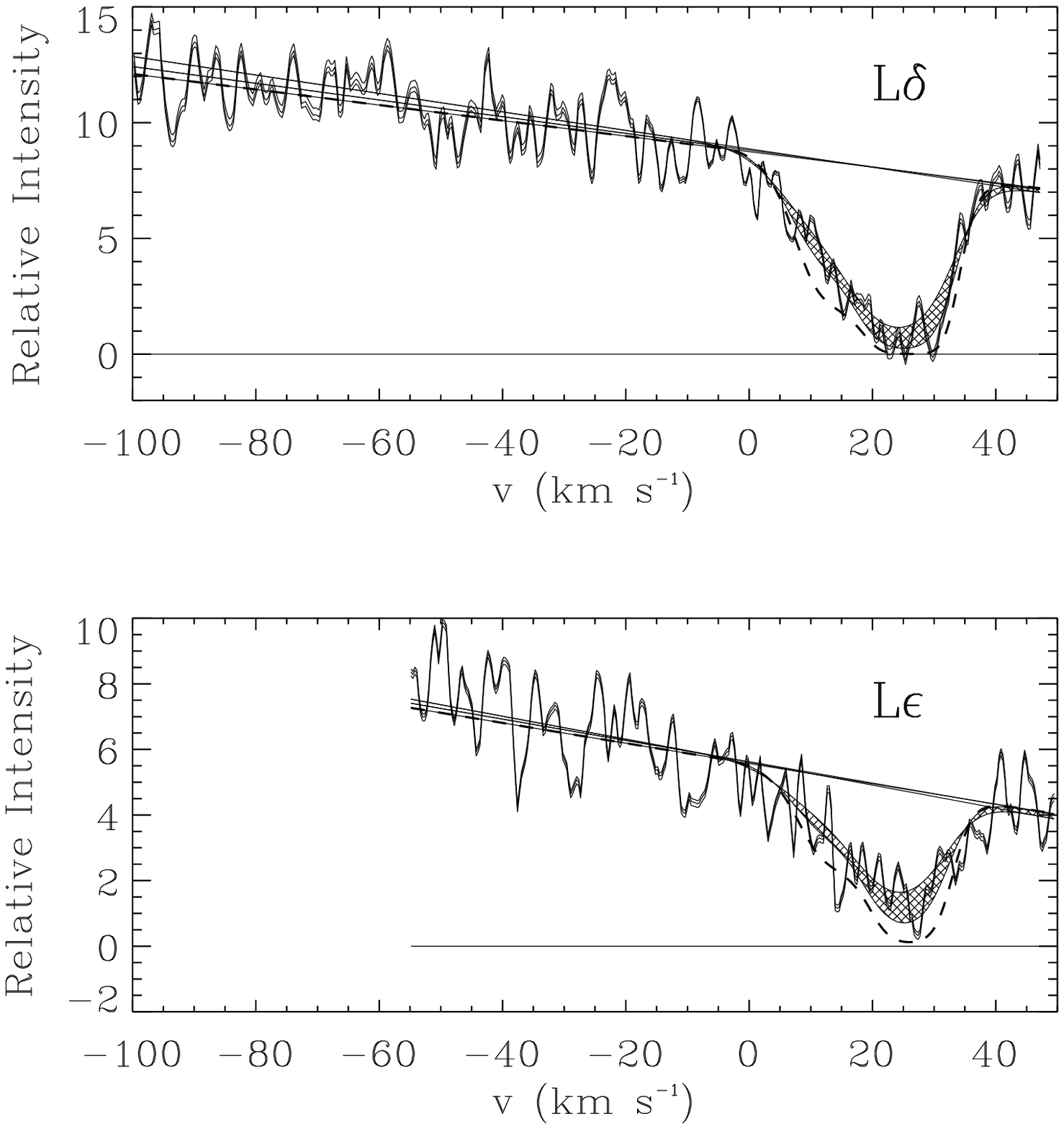}
\caption{Regions containing acceptable model deuterium L$\delta$ and
L$\epsilon$ profiles (crosshatched zones), bounded by the computed
profiles that correspond to the lower and upper bounds of the 90\%
confidence interval for $N$(D~I) given in
Table~\protect\ref{chi2_results} (smooth, solid curves), plotted on top
of the actual data.  Very small changes in the continuum levels sought
by the $\chi^2$ minimizations are also shown for the limiting values [in
both cases, the lower $N$(D~I) seeks a continuum with a more negative
slope].  Very small differences in the favored values of the free
parameters (3) through (6) identified at the end of
\S\protect\ref{errors} lead to small differences in the adjusted fluxes. 
Dashed lines show the profiles that would be expected if $N({\rm
D~I})=2.34\times 10^{15}{\rm cm}^{-2}$, a value that would be consistent
with ${\rm D/H}=1.5\times 10^{-5}$ measured for the local interstellar
medium in front of $\alpha$~Aur \protect\markcite{3085} (Linsky et al.
1995) and HR1099 \protect\markcite{4220} (Piskunov et al. 1997).
\label{extremes}}
\end{figure}

The most important terms in the summation for $\chi^2$ are those that
are directly influenced by trial values of $N$(D~I) and $T$, through
differences over the (deuterium line) velocity interval $-10$ to
$+40\,{\rm km~s}^{-1}$ between measured intensities $I_{\rm meas}$ and
the computed values of $I_{\rm exp}$, the expected absorption profile
multiplied by a local continuum level.  At the same time, parameters
that define the continuum contribute to $\chi^2$ through the deviations
between $I_{\rm meas}$ and straight-line extrapolations over the
intervals ($-100$ to $-10$, +40 to $+47\,{\rm km~s}^{-1}$) for L$\delta$
and ($-55$ to $-10$, +40 to $+50\,{\rm km~s}^{-1}$) for L$\epsilon$ (see
Fig~\ref{extremes}).  Likewise, $\chi^2$ is influenced by modifications
in the background zero level that must be subtracted from the raw
intensities at all velocities: we allowed the sum to include deviations
away from zero for the background-corrected fluxes over the heavily
saturated portion of the H profile from +60 to $+140\,{\rm km~s}^{-1}$
(for the D line heliocentric velocity scales shown in
Figs.~\ref{dorideut} and \ref{extremes}).  Finally, fluxes over all
velocities in the L$\delta$ profile can be modified by the contamination
correction signal, whose amplitude was allowed to vary as we searched
for a minimum $\chi^2$ in each case.

We had a total of 320 independent intensity measurements to constrain
the 10 free parameters listed at the end of \S\ref{errors}, so we should
insist that the minimum $\chi^2$ agree with a reasonable expectation for
${\rm df}=310$.  In fact, with our original estimate for the noise in
the measurements, we arrived at a minimum $\chi^2 = 225$, a value that
was unreasonably low.  In later calculations, we rescaled this noise
level by a factor $\sqrt{225/278}=0.90$, since we had a 90\% confidence
that the minimum $\chi^2$ should be at least equal to 278 for df~=~310. 
We felt that it was legitimate for us to perform a {\it post facto\/}
rescaling of the noise, because our original estimate was accurate to
only a level of about 25\%.  This rescaling is a conservative one,
because it's more probable that the minimum $\chi^2$ should really be
about equal to 310.  If we had used 310 instead of 278 in the expression
for the noise multiplication factor, we accordingly would have found
tighter limits for $N$(D~I) because the $\chi^2$ expressions would have
increased more rapidly as we deviated away from the most probable
$N$(D~I).

To find the minimum $\chi^2$, we used Powell's method of converging to
the minimum of a multi-dimensional function \markcite{3558} (Press et
al. 1992, p. 406).  After finding this minimum and noting the most
probable $N$(D~I), we then evaluated the confidence interval for
$N$(D~I) by forcing this parameter to vary, but at the same time
allowing the other 9 parameters to adjust to new minima in $\chi^2$. 
Our target values for the new minima corresponded to $\chi^2({\rm
min})+4.6$ and $\chi^2({\rm min})+9.2$ for the 90\% and 99\% confidence
limits (i.e., ``1.65$\sigma$'' and ``2.58$\sigma$'' deviations),
respectively, where $\chi^2({\rm min})$ is the overall minimum at the
preferred value of $N$(D~I) as shown in Table~\ref{chi2_results}.  This
exercise ultimately led to the limiting values for $N$(D~I) listed in
the table.  Over the full range of $N$(D~I) between the most extreme
limits, the temperature $T$ was the only parameter that showed any
profound change.  For this reason, $T$ is also listed.  Our result for
the most probable $N$(D~I) is in near perfect agreement with the value
$\log N({\rm D~I})=15.08$ reported by Laurent, et al \markcite{1784}
(1979) in their investigation that led to a value D/H~=~$7\times
10^{-6}$ using data from {\it Copernicus}.

\placetable{chi2_results}
\begin{deluxetable}{lccc}
\tablecaption{Limits for $N$(D~I) from the $\chi^2$ Analysis
\label{chi2_results}}
\tablewidth{0pt}
\tablehead{
\colhead{Significance} & \colhead{$N$(D~I)} &
\colhead{$T$\tablenotemark{a}} & \colhead{$\chi^2$}\\
\colhead{} & \colhead{($10^{15}{\rm cm}^{-2}$)} & \colhead{(K)} &
\colhead{}}
\startdata
Minimum $N$(D~I) at the 99\% confidence limit&0.89&6800&286.1\nl
Minimum $N$(D~I) at the 90\% confidence limit&0.96&6600&281.7\nl
Best $N$(D~I)\dotfill&1.16&5900&277.0\nl
Maximum $N$(D~I) at the 90\% confidence limit&1.45&3500&281.7\nl
Maximum $N$(D~I) at the 99\% confidence limit&1.61&2400&286.2\nl
\enddata
\tablenotetext{a}{The 99\% confidence limits for $T$ if the value of
$N$(D~I) is allowed to float is 1000\,K [at $N$(D~I) = $1.41\times
10^{15}{\rm cm}^{-2}$] and 14,500\,K [at $N$(D~I) = $1.12\times
10^{15}{\rm cm}^{-2}$].  [The 90\% limits are (2200\,K, $1.30\times
10^{15}{\rm cm}^{-2}$) and (11,000\,K, $1.12\times 10^{15}{\rm
cm}^{-2}$).] This indicates that we are not including physically
implausible temperatures in the simultaneously permitted combinations
for the two parameters $N$(D~I) and $T$.  It also reveals that extreme
deviations in $T$ have only a mild effect on our answer for N(D~I),
which is a good justification for our simplifying assumption that $T$
does not vary between superposed velocity components.}
\end{deluxetable}

Fig.~\ref{extremes} shows the observed deuterium profiles along with the
expected absorption profiles (upper and lower boundaries of the
crosshatched regions) whose shapes are determined by the shape of the
N~I profile (heavy, solid line in Fig.~\ref{NandO}) after it has been smoothed
to allow for possible extra thermal Doppler motions that would be
expected for the lighter atoms (dashed line in Fig.~\ref{NandO}). 
Basically, apart from the thermal smearing, there is no evidence that
there are deviations between the nitrogen and deuterium velocity
profiles.

\placefigure{extremes}

In Fig.~\ref{extremes} we also illustrate with a dashed line the depth
and shape of the expected profile if $N({\rm D~I})$ were as high as
$2.34\times 10^{15}{\rm cm}^{-2}$, a value that would give ${\rm
D/H}=1.5\times 10^{-5}$ as seen elsewhere (\S\ref{bkgnd}) if $N({\rm
H~I})=1.56\times 10^{20}{\rm cm}^{-2}$ (\S\ref{N(HI)} below).  When
$N({\rm D~I})$ is forced to this large value, $\chi^2_{\rm min}$ occurs
at $T\leq 300$K.  This value seems unrealistically low, in view of the
evidence that the 0$-$1 rotational temperature of H$_2$ toward
$\delta$~Ori~A is 1625\,K \markcite{1141} (Savage et al. 1977).  Thus,
we set a constraint $T=1625$\,K (but allowed other free parameters to
float) when we constructed the dashed line in Fig.~\ref{extremes}.  For
this case, $\chi^2-\chi^2_{\rm min}=36.3$ which is clearly unacceptable.

\section{A Redetermination of $N$(H~I)}\label{N(HI)}

Published values of $N$(\ion{H}{1}) based on moderate resolution
recordings of the L$\alpha$ absorption in the spectrum of $\delta$~Ori
range from $1.25^{+0.33}_{-0.28}\times 10^{20}\,{\rm cm}^{-2}$
\markcite{1369} (Jenkins 1970), based on photographic spectra recorded
on sounding rocket flights, to $1.7\pm 0.34\times 10^{20}\,{\rm
cm}^{-2}$ \markcite{1026} (Bohlin, Savage, \& Drake 1978) from a
spectrum recorded by {\it Copernicus}.  Since $\delta$~Ori is a hot star
(O9.5II), the stellar \ion{H}{1} absorption line makes a negligible
contribution to $N$(\ion{H}{1}).\footnote{Diplas \& Savage
\protect\markcite{2712} (1994) have estimated that the equivalent H~I
column density caused by stellar absorption for $\delta$~Ori is
10$^{17.6}$ cm$^{-2}$. Therefore correcting for the stellar L$\alpha$
line changes log $N$(H I) by only 0.01~dex.} The accuracies of these
measurements are satisfactory for studies of general trends, but for
measuring D/H, and in particular to investigate the possible spatial
variability of D/H, we must strive for a precision in $N$(\ion{H}{1})
that is as good as or preferably better than that for $N$(\ion{D}{1}). 

The H absorption features shown in Fig.~\ref{dorideut} are of no use in
determining $N$(\ion{H}{1}) because the lines are heavily saturated, and
most of the absorption is by small amounts of hydrogen with velocities
well displaced from the line core.  The damping wings of these lines are
too weak to measure.  In contrast, the L$\alpha$ feature has very strong
wings, ones that make this feature the least susceptible of all the
Lyman series lines to any contributions from high-velocity wisps of
\ion{H}{1} that do not produce detectable counterparts in \ion{D}{1}
absorption.   It is for this reason that we concluded that the L$\alpha$
feature was the best indicator of $N$(H~I).

Spectra of $\delta$~Ori obtained with the {\it International Ultraviolet
Explorer\/} (IUE) in high-dispersion mode (FWHM $\approx$ 25 km
s$^{-1}$) were particularly attractive for our study of the L$\alpha$
feature for several reasons. First, the alternative was to use archival
{\it Copernicus\/} data  (L$\alpha$ was not recorded by IMAPS or HST),
but the only available {\it Copernicus\/} data with sufficiently broad
wavelength coverage of L$\alpha$ were obtained with the low-resolution
U2 detector, and for this detector uncertainties are introduced by stray
light from the vent hole \markcite{1393} (Rogerson et al. 1973), an
effect that requires a special correction \markcite{1771} (Bohlin 1975)
of uncertain accuracy. Second, a large number of observations of
$\delta$~Ori obtained under slightly different observing conditions
(e.g., small aperture {\it vs.\/} large aperture) over the course of
many years are available from the IUE archive. By analyzing all of the
IUE exposures rather than a single observation, we can validate our
estimates of random errors and also increase our chances of exposing
{\it systematic\/} errors in the derivation of $N$(\ion{H}{1}). 
Finally, our ability to combine many observations allowed us to reduce
the effects from random noise by brute force.

A potentially important source of systematic error is the fact that
$\delta$~Ori is a complex multiple-star system.  The primary (component
A) is a single-lined spectroscopic binary that is important in the
history of ISM research: stationary Ca~II absorption features in its
spectrum provided the earliest evidence of interstellar gas
\markcite{2084} (Hartmann 1904).  The velocity amplitude of the binary
is $98\,{\rm km~s}^{-1}$, and it has a period of 5.7 days
\markcite{3597} (Harvey et al. 1987).  It is also a partially eclipsing
binary \markcite{3594} (Koch \& Hrivnak 1981), and it has a visual
companion of comparable brightness at a separation of $\sim 0\farcs 2$
\markcite{3612, 3595} (Heintz 1980; McAlister \& Hendry 1982). The
spectroscopic binary nature of the star system presents an opportunity
to investigate systematic errors in the determination of the
interstellar $N$(\ion{H}{1}): the \ion{H}{1} L$\alpha$ profile is
extremely broad, and if there are unrecognized stellar lines in the
principal part of the Lorentz wings of the L$\alpha$ feature, then their
additional optical depth could lead to an overestimate of
$N$(\ion{H}{1}). However, such stellar lines should move in velocity as
the binary traverses its orbit, and this may lead to different values of
$N$(\ion{H}{1}) when observations made at different times are analyzed.
If this occurs, then we should see $N$(\ion{H}{1}) change as a function
of the spectroscopic binary phase. Similarly, we can check for
systematic changes in $N$(\ion{H}{1}) when the multiple star enters the
partial eclipses over the phase intervals 0.9$-$0.1 and 0.4$-$0.6. 
While any dependence on phase may uncover an influence that stellar
features have on the outcome, there is no guarantee that they do not
perturb our result in a manner that is uniform over all phases.

According to the NSSDC archive, there are 59 IUE observations of
$\delta$~Ori obtained with the Short Wavelength Prime (SWP) camera in
the high-dispersion echelle mode. We screened these observations for
saturated exposures, missing data, or other problems and rejected two of
the observations, leaving 57 spectra for our analysis. Using the
standard IUE RDAF software, we selected the spectral regions of interest
from the standard IUESIPS data rather than the NEWSIPS reductions since
there are a number of problems with NEWSIPS processing as applied to
high dispersion spectra that could adversely affect our analysis
\markcite{3609} (Massa et al. 1998).

In the vicinity of the L$\alpha$ line the orders on the IUE detector are
closely spaced, and scattered light from adjacent orders overlaps in the
interorder region causing an incorrect background subtraction and zero
intensity level when using the standard software. We used the method of
Bianchi \& Bohlin \markcite{3596} (1984) to correct for this problem.
Also, in some cases, a velocity shift was applied to the IUE data based
on the position of the \ion{N}{1} triplet at 1200\,\AA\ compared to that
expected from the IMAPS \ion{N}{1} profile (\S\ref{template}) with
optical depths rescaled to account for the stronger transition
probabilities. This should register all of the IUE data to the correct
velocity scale to an accuracy of better than $\pm$5 km s$^{-1}$, which
is more than adequate for a determination of $N$(\ion{H}{1}) since we
are fitting {\it both\/} of the strong damping wings of the L$\alpha$
profile (an error of 5 km s$^{-1}$ in the velocity scale zero point
changes $N$(\ion{H}{1}) by an insignificant amount).

IUE data contain a number of perturbations in addition to the usual
photon counting noise. They show strong fixed pattern noise, ``hot
spots'' which mimic emission features, and drop-outs where the reseaux
used to correct for camera distortions happen to fall on the spectrum
\markcite{3598} (Harris \& Sonneborn 1987). In addition, the spectra
occasionally show artifacts at the transitions between echelle orders
due to errors in the ripple correction (see below). Again, by analyzing
{\it all\/} of the IUE data, we can reduce the impact of these noise
sources, which are present in some of the observations and are not
apparent in others.  We ascertained that there was no {\it persistent}
nonlinearity in the photometric response of IUE by comparing its
L$\alpha$ profiles of HD\,93521 and HD\,74455 with those recorded for
the same stars by the GHRS on HST with the medium-resolution grating
(G160M).  Departures from the GHRS spectra near the breaks in the IUE
echelle orders seem to come and go, but aside from the greater random
noise in the IUE spectra, the spectra are usually very similar to each
other.

In simplest terms, our means for constraining the \ion{H}{1} column
density followed a method introduced by Jenkins \markcite{1312} (1971)
and used later by Bohlin \markcite{1771} (1975), Bohlin, et al
\markcite{1026} (1978), Shull \& van Steenberg \markcite{1058} (1985),
and Diplas \& Savage \markcite{2712} (1994): we determined the
$N$(\ion{H}{1}) that provides the best fit to the L$\alpha$ profile with
the optical depth $\tau$ at a given wavelength $\lambda$ calculated from
the expression
\begin{equation}
\tau (\lambda ) = N({\rm H~I})\sigma (\lambda) = 4.26\times 10^{-20}
N({\rm H~I})(\lambda - \lambda _{0})^{-2}
\end{equation}
\markcite{1312} (Jenkins 1971), where $\lambda_0$ is the L$\alpha$ line
center at the velocity centroid of the hydrogen. However, we went a step
further by employing the technique used to estimate $N$(\ion{D}{1}) in
\S\ref{N(DI)}, i.e., we first determined the important free parameters
that could be adjusted to fit the \ion{H}{1} L$\alpha$ absorption
profile, then we found the set of parameters that minimized $\chi ^{2}$
using Powell's method, and finally we set confidence limits on the
\ion{H}{1} column density by increasing (or decreasing) $N$(\ion{H}{1})
with the other parameters freely varying until $\chi ^{2}$ increased by
the appropriate amount for the confidence limit of interest.

\begin{figure}
\plotfiddle{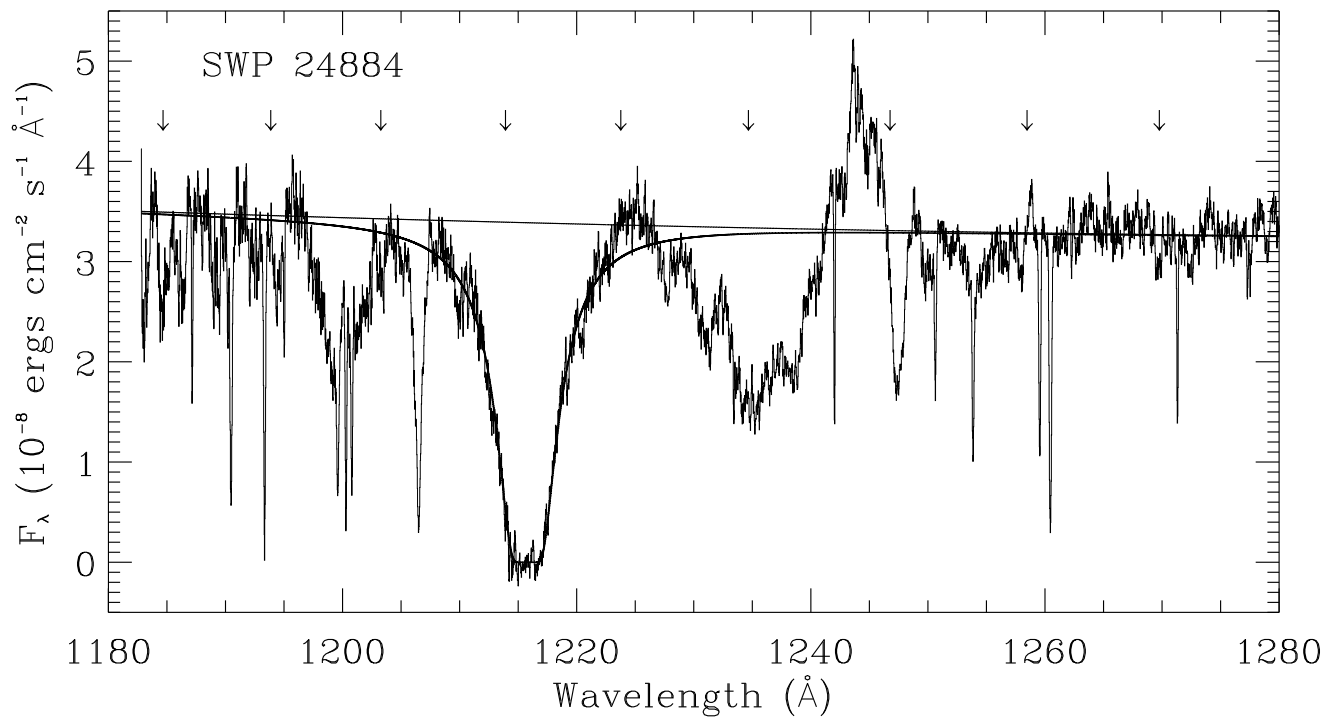}{4in}{90}{75}{75}{300}{0}
\caption{A sample high-dispersion spectrum of $\delta$~Ori~A in the
vicinity of L$\alpha$ showing the second-order polynomial fitted to the
continuum and the computed profile for the preferred value of
$N$(\ion{H}{1}).  This IUE spectrum has been smoothed with a 5-pixel
boxcar for display purposes only; the unsmoothed data were used to
constrain $N$(\ion{H}{1}) as described in the text. Due to the presence
of the \ion{N}{5} P-Cygni profile longward of L$\alpha$ and several
strong lines (e.g., \ion{Si}{3} $\lambda$1206.5) shortward, the
continuum fitting regions are somewhat far-removed from the L$\alpha$
region. Nevertheless, the continuum is well constrained. The transitions
between orders are indicated with arrows. Small artifacts are
occasionally present at some order transitions due to errors in the
ripple correction.\label{hicon}}
\end{figure}

\placefigure{hicon}

Figure~\ref{hicon} shows a sample IUE spectrum in the vicinity of the
L$\alpha$ absorption line. From this figure one can see that the
continuum is close to linear in this region. However, it is possible
that the continuum has a slight downward or upward curvature, so we
assumed a second-order polynomial to describe the continuum and allowed
the $\chi ^{2}$ minimization process to determine the continuum shape
that provided the best fit to the L$\alpha$ profile. Therefore the free
parameters for fitting the \ion{H}{1} profile were $N$(\ion{H}{1}),
three coefficients that specify the second-order continuum polynomial,
and a simple additive correction to the intensity zero
point.\footnote{Despite our use of the Bianchi \& Bohlin
\protect\markcite{3596} (1984) correction, in many cases inspection of
the flat-bottomed, saturated portion of the L$\alpha$ profile showed
that the zero intensity level was not quite correct, so we included a
zero point shift as a free parameter and used intensity points within
the saturated core as one of the collections of terms for calculating
$\chi ^{2}$.}

We point out that the continuum placement is constrained not only by the
fits to regions that are far removed from the L$\alpha$ feature, but
also by requiring a good match to the shapes of its damping wings.  For
instance, if the continuum is badly placed or has too much upward or
downward curvature, then a poor fit results.  In particular, if we
artificially forced the continuum to have a downward curvature (in an
experimental challenge to lower $N$(H~I) and thus provide a higher D/H),
we obtained clearly inferior fits.  We found that in the course of our
minimizing $\chi^2$ that we could always simultaneously obtain a good
fit to the profile and match the outlying fluxes with a nearly flat
continuum.

\begin{figure}
\plotone{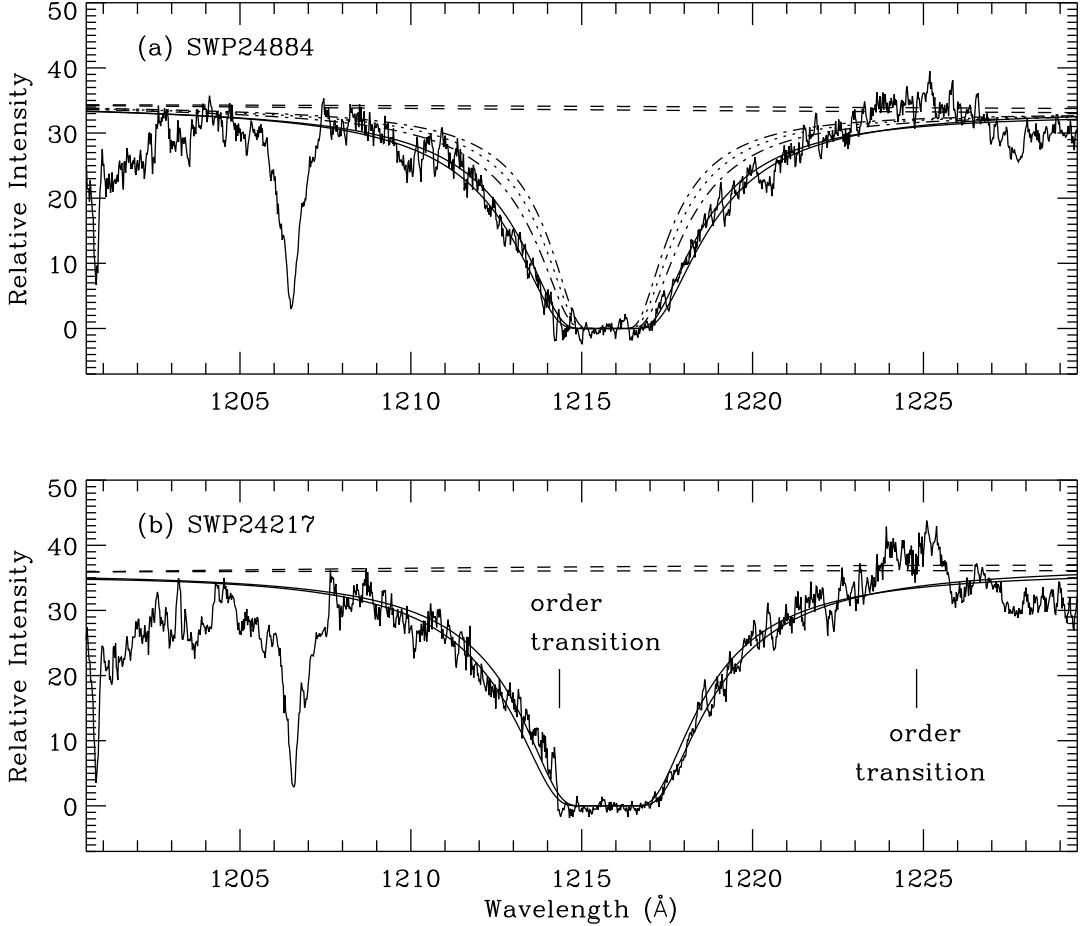}
\caption{Examples of high-dispersion IUE spectra of the \ion{H}{1}
L$\alpha$ line in the spectrum of $\delta$~Ori~A. Panel (a) shows a
normal IUE spectrum while (b) shows a spectrum with the peculiar
features at the order transitions near 1214.4 and 1224.8\,\AA. Computed
profiles for the upper and lower bounds on $N$(\ion{H}{1}) at the 90\%
confidence levels for the individual cases are overplotted on the IUE
data, and the continua corresponding to these upper and lower bounds are
shown with dashed lines.  The narrower damping profiles well inside the
observed profile in the upper panel indicate the expected appearance if
D/H were equal to the more general result $1.5\times 10^{-5}$ discussed
in \S\protect\ref{bkgnd} with our values of $N$(D~I) given in
Table~\protect\ref{chi2_results} (dotted line = most probable value,
dot-dashed line = 90\% confidence limits).  The IUE spectra have been
smoothed with a 5-pixel boxcar for display purposes only; the unsmoothed
data were used to constrain $N$(\ion{H}{1}) as described in the text.
\label{hiprofs}}
\end{figure}

\placefigure{hiprofs}

Figure ~\ref{hiprofs} shows two examples of \ion{H}{1} L$\alpha$
profiles observed with IUE, along with the computed profiles for the
lower and upper bounds on $N$(\ion{H}{1}) at the 90\% confidence level
for each case.  The final continua corresponding to the upper and lower
bounds are plotted with dashed lines.  Panel ($a$) shows a typical
spectrum, while ($b$) shows examples of artifacts at $\lambda \ \sim$
1214.4 and 1224.8\,\AA\ that we encountered at the transitions between
IUE echelle orders. Assuming these to be artifacts due to the ripple
correction, we used only the sides of the L$\alpha$ profile in the
wavelength ranges 1209.0$-$1213.5\,\AA\ and 1217.14$-$1223.0\,\AA\ and
thereby excluded these artifacts from the $\chi^{2}$ calculation.  This
procedure resulted in bounding profiles such as those shown in both of
the panels of the Figure. However, as an experiment we also processed
all of the IUE data including the region at $\sim$ 1214.4\,\AA\ in the
$\chi^{2}$ calculation in order to evaluate the importance of this
effect on the final results (see below).

It is important to note that the great strength of the L$\alpha$ profile
makes the Lorentzian wings dominate over the effects of instrumental or
Doppler broadening.  Gas that is known to exist in the vicinity of the
Orion association at high velocities ($v\approx -100\,{\rm km~s}^{-1}$)
\markcite{1151} (Cowie, Songaila, \& York 1979) should not be important,
since the absorptions from any wisps of H~I at such velocities are
displaced by only about 0.3\,\AA\ relative to the line center.

\begin{figure}
\plotone{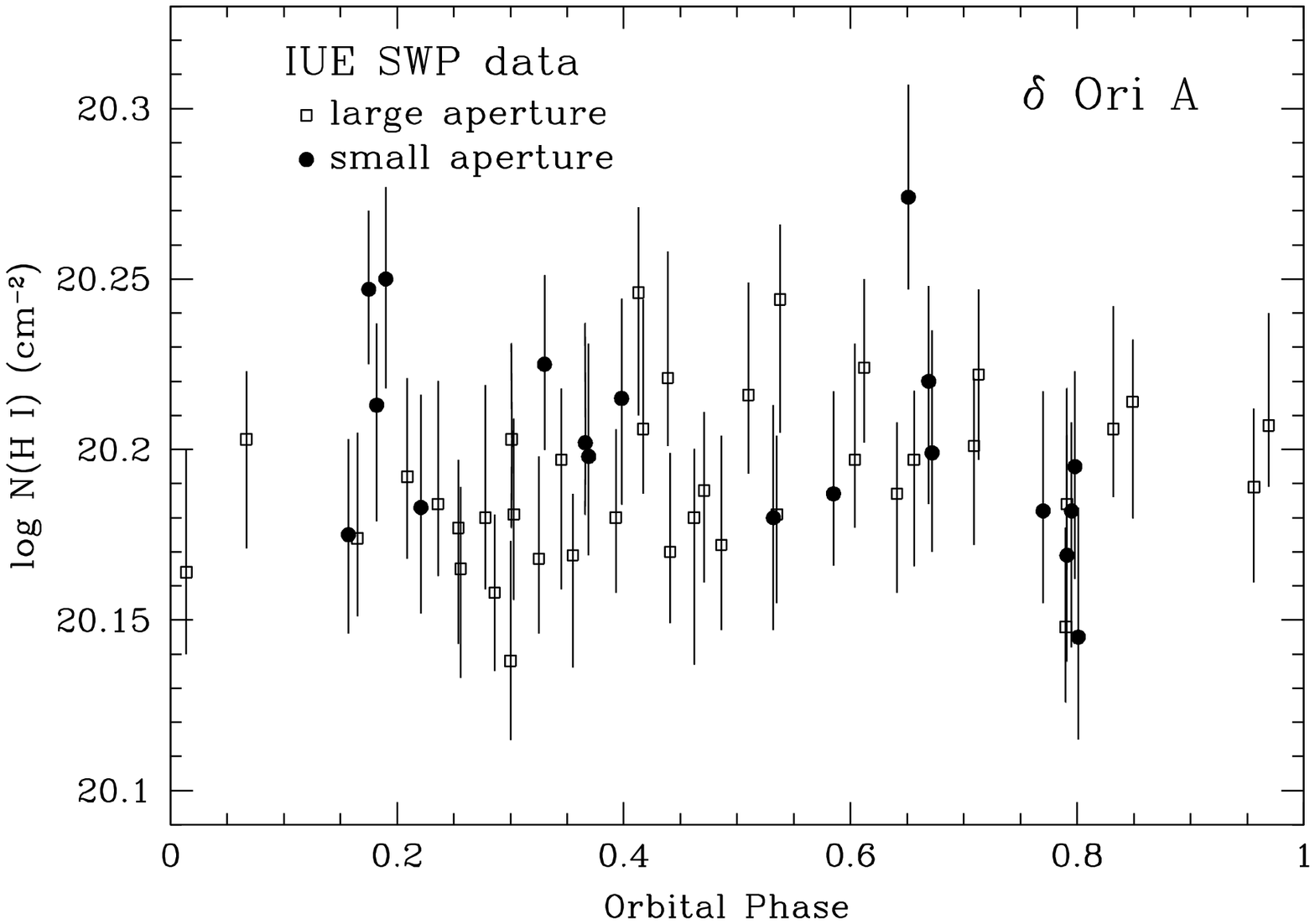}
\caption{\ion{H}{1} column densities toward $\delta$~Ori~A derived from
57 IUE observations as described in the text, plotted versus phase of
the spectroscopic binary. Each column density is plotted with $\pm
1\sigma$ error bars. Large aperture data are shown with open squares
while small aperture data are indicated with filled
circles.\label{allhi}}
\end{figure}

\placefigure{allhi}

Figure ~\ref{allhi} shows the logarithms of the derived \ion{H}{1}
column densities with their 1$\sigma$ error bars, plotted as a function
of the spectroscopic binary phase, for all of the IUE SWP data except
the two rejected exposures. We calculated the phase using the period and
$T_{0}$ derived by Harvey, et al. \markcite{3597} (1987) from their
analysis of all suitable data from 1902$-$1982 (including some of the
IUE data used here). There are no obvious systematic trends as function
of spectroscopic binary phase evident in this plot, which gives us some
assurance that weak stellar lines do not significantly affect the
derived $N$(\ion{H}{1}). In the figure, large and small aperture data
are shown with different symbols to check for any systematic
differences, and no differences are readily apparent. With only the
large aperture data we derive a mean \ion{H}{1} column density of
$<N$(\ion{H}{1})$>$ = 1.54$\times 10^{20}$ cm$^{-2}$ with an rms
dispersion $\sigma$ = $0.08\times 10^{20}$ [both quantities are weighted
inversely by the variances of the individual $N$(\ion{H}{1}) estimates].
With the small aperture data, we obtain $<N$(\ion{H}{1})$>$ =
1.59$\times 10^{20}$ cm$^{-2}$ with $\sigma$ = $0.11\times 10^{20}$.
Therefore it appears appropriate to combine the large and small aperture
data to constrain $N$(\ion{H}{1}). Using the entire IUE data set, we
obtain $<N$(\ion{H}{1})$>$ = 1.56$\times 10^{20}$ cm$^{-2}$ with an rms
scatter equal to $0.09\times 10^{20}$. Since there are 57 measurements,
the error in the mean = $\sigma/\sqrt{57}$ = 0.01$\times 10^{20}$ for
the whole data set.  We find that including the region that spans the
ripple correction artifact at 1214.4\,\AA\ shown in Figure
~\ref{hiprofs}(b) lowers the overall result by less than 0.01~dex.  This
is because the feature is present in only a small fraction of the IUE
spectra.

The scatter in Figure ~\ref{allhi} appears to be due entirely to the
uncertainties from noise in the individual measurements: the value for
$\chi ^{2}=\sum_i \left\{ \left[ N_i({\rm H~I})-\langle N({\rm
H~I})\rangle\right]/\sigma\left[ N({\rm H~I})\right]_i \right\}^2$
calculated for the entire data set is 50.32, which implies a reduced
$\chi^2$, $\chi ^{2}/56 = 0.90$.  {\it Nevertheless, given the many
potential sources of systematic error in these particular IUE data, it
is still possible that there are some unrecognized systematic errors
which affect N(H~I) and that the real error in the mean is
underestimated.}  However, the good fits to the \ion{H}{1} L$\alpha$
profiles (see Fig.~\ref{hiprofs}) and the lack of pronounced variations
with binary phase indicate that such unrecognized systematics are not
likely to be large.\footnote{In their study of an observation of
L$\alpha$ absorption in the spectrum $\mu$~Col, Howk, Savage and Fabian
\markcite{HSF99}(1999) estimated a systematic error of 0.02~dex for a
determination of $N({\rm H~I})$ somewhat less than $10^{20}{\rm cm}^{-2}$.
This relative error is about half of the amount that would be
needed to have any appreciable impact on the relative errors for our D/H
toward $\delta$~Ori~A given in \S\ref{D/H}.  Although we must
acknowledge that their GHRS spectrum is of much better quality than any
that were taken by IUE, their independent estimate that gives a low
value for the magnitude of a systematic error in this type of
measurement is encouraging.}  It is reassuring to note that a constant,
systematic error in the result for $N({\rm H~I})$ would need to be at
least 15 times as large as the formal (random) error before it could
have a meaningful effect on the overall error for D/H derived in
\S\ref{D/H} below.  In the light of our deliberate attempt to uncover
such systematic errors in the study of $N({\rm H~I})$ {\it vs.\/}
orbital phase, we are confident that they could not have such a large
numerical advantage, which means they are unlikely to be a critical
issue in this investigation.

Finally, to illustrate more graphically our confidence in the H~I
result, we show with dotted and dash-dot damping profiles in
Fig.~\ref{hiprofs}(a) the expected appearance of the L$\alpha$
absorption if our $N$(D~I) derivations are correct but $N$(H~I) were low
enough to make ${\rm D/H}=1.5\times 10^{-5}$, as indicated for other
lines of sight (\S\ref{bkgnd}).  There seems to be no question that the
real data are strikingly inconsistent with these lower values for
$N$(H~I).

\section{D/H toward $\delta$~Ori and its Significance}\label{D/H}

Combining the results reported in \S\ref{N(DI)} and \S\ref{N(HI)}, we
find that with 90\% confidence we can declare that $N({\rm D~I})/N({\rm
H~I})= 7.4^{+1.9}_{-1.3}\times 10^{-6}$ in the direction of
$\delta$~Ori~A.  The most noteworthy feature of this result is that it
differs from most determinations of D/H along other lines of sight in
our local region of the Galaxy (\S\ref{bkgnd}).   It is clear that our
value for D/H represents a deviation, even if one relies only on the HST
observations that generally concentrate within the range ${\rm D/H} =
1.3-1.7\times 10^{-5}$ in the very local medium and rejects the {\it
Copernicus\/} results because their accuracies may have been overstated. 
Our result shows a simple velocity structure for the D~I, N~I and O~I
absorption profiles and thus removes the primary uncertainty that
confronted Laurent, et al \markcite{1784} (1979).  It also removes the
grounds on which McCullough \markcite{178} (1992)
rejected the measurements of deuterium toward stars in Orion.

\section{Abundances of Heavier Elements}\label{heavy}

Variations in the heavy element abundances of stars with similar ages,
such as B stars in the Orion association \markcite{3500} (Cunha \&
Lambert 1994) or F and G type stars at a Galactocentric radius nearly
the same as the Sun \markcite{3421}  (Edvardsson et al. 1993), suggest
that the ISM out of which the stars formed may be a heterogeneous
mixture of gases with different levels of heavy element enrichment. 
This may possibly result from random dilutions of gas in the Galactic
plane by metal-poor material falling in from the halo \markcite{2858}
(Meyer et al. 1994).  If the material in the direction of $\delta$~Ori
had a lower deuterium abundance because it had been subjected to more
intensive stellar processing or less of this dilution, we would expect
the heavy element abundances to be higher than elsewhere.  For our test
of this proposition, we will examine the abundances of oxygen and
nitrogen, both of which are unlikely to be appreciably altered by
depletions onto dust grains \markcite{4196, 3466} (Meyer, Cardelli, \&
Sofia 1997; Meyer, Jura, \& Cardelli 1998).  O~I and N~I are also good
standards because their ionizations are closely coupled to that of H~I
(\S\ref{template}), and this allows us to neglect the higher stages of
ionization because they should be identified only with ionized H and D. 
O and N are also useful for comparisons with abundance studies elsewhere
in the Universe.  These investigations are facilitated by the generous
number of transitions in the ultraviolet with a broad range of $f$
values \markcite{4101} (Timmes et al. 1997).  [Argon is another element
that is expected to have very little depletion, but it is not suitable
for comparison because under many circumstances its ionization can
differ appreciably from that of H \markcite{4310} (Sofia \& Jenkins
1998).]

Meyer, et al. \markcite{4196} (1998) found from their measurement of the
intersystem O~I transition at 1356\,\AA\ that toward $\delta$~Ori ${\rm
O/H}=2.82\pm0.46\times 10^{-4}$ (from the same data that we used to
construct the profile shown in Fig.~\ref{NandO}).  This
value\footnote{The value for $N$(H~I) adopted by Meyer, et al.
\protect\markcite{4196} (1998) was $1.6\times 10^{20}{\rm cm}^{-2}$
which is very close to the result $1.56\times 10^{20}{\rm cm}^{-2}$ that
we derived in our more precise analysis (\S\protect\ref{N(HI)}).} is
slightly less than their average of $3.19\pm 0.14\times 10^{-4}$ over 13
lines of sight.  Within the experimental errors, however, the value is
consistent with the average.  Meyer, et al. \markcite{3466} (1997) found
that ${\rm N/H}=7.5\pm0.4\times 10^{-5}$ toward 6 stars.  Since
$\delta$~Ori was not part of this sample, we must rely on our own
measurement of N~I which yields $\int N_a(v)dv = 6.2\times 10^{15}{\rm
cm}^{-2}$ -- see \S\ref{template} and the N~I profile in
Fig.~\ref{NandO}.  With our result for $N$(H~I), we arrive at ${\rm
N/H}=4.0\times 10^{-5}$. Thus, we see no evidence that the gas has been
specially enriched with material having a high abundance of heavy
elements and, by virtue of a more intensive exposure to stellar
interiors, a more thorough depletion of deuterium.  In fact, the
possible positive correspondence in the deviations in the N and D
abundances is reminiscent of the suggested correlation (but with large
errors) between D/H and Zn/H shown by York \& Jura \markcite{1040}
(1982).  Of course, we must acknowledge that for nitrogen a comparison
of our result for $\delta$~Ori with the general measurements of Meyer,
et al \markcite{3466} (1997) for other lines of sight may be compromised
by errors in f values, since we used different transitions to determine
$N({\rm N~I})$.

\section{Discussion}\label{disc}

Our finding presented in \S\ref{D/H} indicates that the most probable
D/H toward $\delta$~Ori is about half as large as that found from
various HST investigations of the local ISM.  Our result applies to an
average over a range of velocities, which means that it represents a
{\it lower limit\/} for the magnitude of deviations from the other
cases.  This anomaly is not linked with an increase in O/H or N/H, as
one might expect from a simple explanation that a greater fraction of
the gas had been cycled through stellar interiors.  Of course, it may be
possible to envision that the gas toward $\delta$~Ori holds an unusually
large fraction of material that has cycled only through the {\it outer
envelopes} of stars, thus depleting the D without increasing the
concentrations of heavier elements.

Recent observations of HD emission in the infrared from gas near
Orion seem to confirm our finding that the abundance of deuterium is low
in this region.  Wright, et al. \markcite{W99}(1999) detected emission
from the HD $J=1\rightarrow 0$ transition at $112\, \mu{\rm m}$ toward
the Orion Bar with the Long Wavelength Spectrometer on board the {\it
Infrared Space Observatory} (ISO).  While there are some uncertainties
in the rotation temperature of HD and the correction factors that must
be applied to observations of the accompanying H$_2$, they arrived at a
preferred value ${\rm HD/H_2}=2.0 \pm 0.6\times 10^{-5}$ which leads to
${\rm D/H}=1.0 \pm 0.3\times 10^{-5}$ since there should be no
appreciable D or H in atomic form.  Their total range for D/H could be as
large as 0.35 to $1.30\times 10^{-5}$ however.  Bertoldi, et al.
\markcite{B99}(1999) used the Short Wavelength Spectrometer on ISO and
found a weak emission from the $J=6\rightarrow 5$ transition at $19.4\,
\mu{\rm m}$ for HD in the Orion molecular outflow OMC-1.  Again,
corrections using information from models of the gas had to be made to
interpret the results.  Bertoldi, et al.\markcite{B99}(1999) concluded
that ${\rm D/H}=7.6\pm 2.9\times 10^{-6}$.  The measurements of HD by
both groups seem to lead to results that are consistent with our
determination of atomic D/H toward $\delta$~Ori~A.

Very distant gas systems that are registered in quasar absorption line
spectra reveal apparent values of D/H that range from $3-4\times
10^{-5}$ \markcite{4365,3643} (Burles \& Tytler 1998a, b) to $\sim
2\times 10^{-4}$ \markcite{2647, 346, 345, 3641, 3632, 3644} (Songaila
et al. 1994; Carswell et al. 1996; Rugers \& Hogan 1996; Wampler et al.
1996; Webb et al. 1997; Tytler et al. 1999)  -- see reviews by Burles \&
Tytler \markcite{3630} (1998c) and Hogan \markcite{3631} (1998).  The
large dispersion in these outcomes might be attributable either to
complications that arise from our incomplete knowledge of the chemical
evolution of systems at large redshifts, the difficulty in obtaining
accurate values of $N$(H~I), or to the presence of random, weak, H~I
systems that masquerade as deuterium by having a velocity offset that is
equal to about $-80\,{\rm km~s}^{-1}$ from a main system.  One might
suppose that, in time, additional observations that include new cases or
new data on existing ones may lead to a better understanding of the
behavior of D/H in the Universe.  Unfortunately, this optimistic belief
may have been dealt a setback by our result that indicates that D/H
could be driven by a process that we do not understand.  Essentially, we
see evidence that the ratio changes over a distance scale where the
environment should be homogeneous, according to observations of other
elements and generally accepted simple models for a galaxy's chemical
evolution and mixing rates in the ISM.

A few proposals to explain possible deviations in the balance of atomic
D to H in the interstellar medium have been considered in the past.  The
simplest involves the selective incorporation of D into HD, an effect
that can amplify HD/H$_2$ to values well above the fundamental ratio of
D to H \markcite{2263} (Watson 1973),
but one that is probably counterbalanced by the more rapid
photodissociation of HD in diffuse clouds because there is no self
shielding (as there often is with H$_2$).  A preferential formation of
HD is not responsible
for the depletion of atomic D toward $\delta$~Ori~A, since $\log N({\rm
HD})<12.8$ \markcite{1015} (Spitzer, Cochran, \& Hirshfeld 1974)
(likewise, we see no HD features in our IMAPS spectrum).  Another
alternative, one advanced by Vidal-Madjar, et al. \markcite{1900} (1978)
and Bruston et al. \markcite{1904} (1981), makes use of the differences
in the ways that D and H can respond to radiation pressure, as a result
of the very different opacities in the Lyman lines.  They proposed that
this effect could lead to a separation of the two species if there were
a density gradient and a nonisotropic radiation field.  Finally, Jura
\markcite{3520} (1982) has suggested that deuterium atoms could collide
with dust grains, stick to them, and then be more strongly bound than
hydrogen.  Furthermore, he suggests that the mobility of the D atoms on
the surfaces of these grains could be much lower than that of H, thus
limiting the chances for combining with H atoms and being ejected as HD. 
One possible way to investigate the plausibility of this hypothesis
might be to look for D$-$C or D$-$O stretch mode absorption features in
dense clouds. 

We look forward to the possibility that insights on the relationship for
the variability in D/H to other interstellar parameters could arise from
the anticipated large increase of information that should come from the
{\it Far Ultraviolet Spectroscopic Explorer\/} (FUSE) after its launch
in mid-1999.  For the local ISM, the FUSE Principal Investigator Team
has identified as potential targets\footnote{Details are given in the
NASA Research Announcement for FUSE, dated Feb 9, 1998 (NRA 98-OSS-02),
or else see http://fusewww.gsfc.nasa.gov/fuse/.} 7 cool stars, 19 white
dwarfs, 9 late B- or early A-type stars, and 7 central stars of
planetary nebulae for studying D/H in the first two years of operations.
\acknowledgments

The ORFEUS-SPAS project was a joint undertaking of the US and German
space agencies, NASA and DARA.  The successful execution of our
observations was the product of efforts over many years by engineering
teams at Princeton University Observatory, Ball Aerospace Systems Group
(the industrial subcontractor for the IMAPS instrument) and Daimler-Benz
Aerospace (the German firm that built the ASTRO-SPAS spacecraft and
conducted mission operations).  Contributions to the success of IMAPS
also came from the generous efforts by many members of the Optics Branch
of the NASA Goddard Space Flight Center (grating coatings and testing)
and from O.~H.~W.~Siegmund and S.~R.~Jelinsky at the Berkeley Space
Sciences Laboratory (deposition of the photocathode material).  This
research was supported by NASA grants NAG5$-$616 to Princeton University
and NAG5$-$3539 to Villanova University.  We thank K.~R.~Sembach for
supplying an IDL routine to apply the Bianchi \& Bohlin correction to
the IUE data.  We also thank A.~Vidal-Madjar, B.~T.~Draine and
B.~D.~Savage for their helpful comments on early drafts of this paper.
The O~I absorption feature at 1355\,\AA\ was observed by the NASA/ESA
Hubble Space Telescope.  This spectral segment was obtained from the
data archive at the Space Telescope Science Institute, operated by AURA
under NASA contract NAS5-26555.  The IUE data were obtained from the
National Space Science Data Center (NSSDC) at NASA's Goddard Space
Flight Center.

\end{document}